\documentclass[pra,aps,showpacs,showkeys,groupedaddress,superscriptaddress,twocolumn,longbibliography]{revtex4-2}

\usepackage{amsmath}
\usepackage{amssymb}  
\usepackage{bbold}    
\usepackage{braket}
\usepackage{graphicx}
\usepackage{subcaption}
\usepackage{fix-cm}
\usepackage{pgfplots} 
\pgfplotsset{compat=1.5} 
\usepackage{color}
\usepackage{float}
\usepackage{ulem}
\usepackage[colorlinks=true,linkcolor=blue,citecolor=blue,urlcolor=blue]{hyperref}  
\usepackage{todonotes}

\usepackage{array}

\usepackage{bm}	

\begin{document}
\title{Quantum theory of fractional topological pumping of lattice solitons}

\author{Julius Bohm}
\affiliation{Department of Physics and Research Center OPTIMAS, RPTU University Kaiserslautern-Landau, 67663 Kaiserslautern, Germany}
\author{Hugo Gerlitz}
\affiliation{Department of Physics and Research Center OPTIMAS, RPTU University Kaiserslautern-Landau, 67663 Kaiserslautern, Germany}
\author{Christina J\"org}
\affiliation{Department of Physics and Research Center OPTIMAS, RPTU University Kaiserslautern-Landau, 67663 Kaiserslautern, Germany}
\affiliation{Research Center QC-AI, RPTU University Kaiserslautern-Landau, 67663 Kaiserslautern, Germany}
\author{Michael Fleischhauer}
\affiliation{Department of Physics and Research Center OPTIMAS, RPTU University Kaiserslautern-Landau, 67663 Kaiserslautern, Germany}
\affiliation{Research Center QC-AI, RPTU University Kaiserslautern-Landau, 67663 Kaiserslautern, Germany}


\begin{abstract}
One of the hallmarks of topological systems
is the robust quantization of particle transport. It is the origin of the integer-valued quantum Hall conductivity and a potential tool for quantum information technology. 
Recent experiments on topological pumps constructed by using 
 arrays of photonic waveguides and described by the (lattice-translational invariant) Aubry-Andr\'e-Harper (AAH) model, have demonstrated 
both integer and fractional transport of lattice solitons. 
In these systems, a background medium mediates interactions between photons via a Kerr nonlinearity and leads to the formation of self-bound multi-photon states.
Upon increasing the interaction strength 
a sequence of transitions was observed from a phase with integer transport in a pump cycle through different phases of fractional transport to a phase with no transport.
We here  present a quantum description of topological pumps of self-bound many-particle states in terms of an effective Hamiltonian of their center-of-mass (COM) motion, which allows to 
introduce an effective band structure $E_\mu(K)$ with $K$ being the COM momentum, and to
classify topological phases in terms of generalized symmetries. We provide an explicit analytic expression of the effective Hamiltonian for few particles in the strong interaction limit and present numerical results in the more general case.
We identify
a topological invariant, an effective single-particle Chern number, 
which fully governs the soliton transport.
Increasing the interaction strength in the AAH model leads to a successive merging of COM bands, 
which is the origin of the observed sequence of topological phase transitions 
and also the potential breakdown of topological quantization for some interaction strength. 
\end{abstract}

\date{\today}
\maketitle


\section{Introduction}

\begin{figure}[H]
    \begin{centering}
        \includegraphics[width=0.9\linewidth]{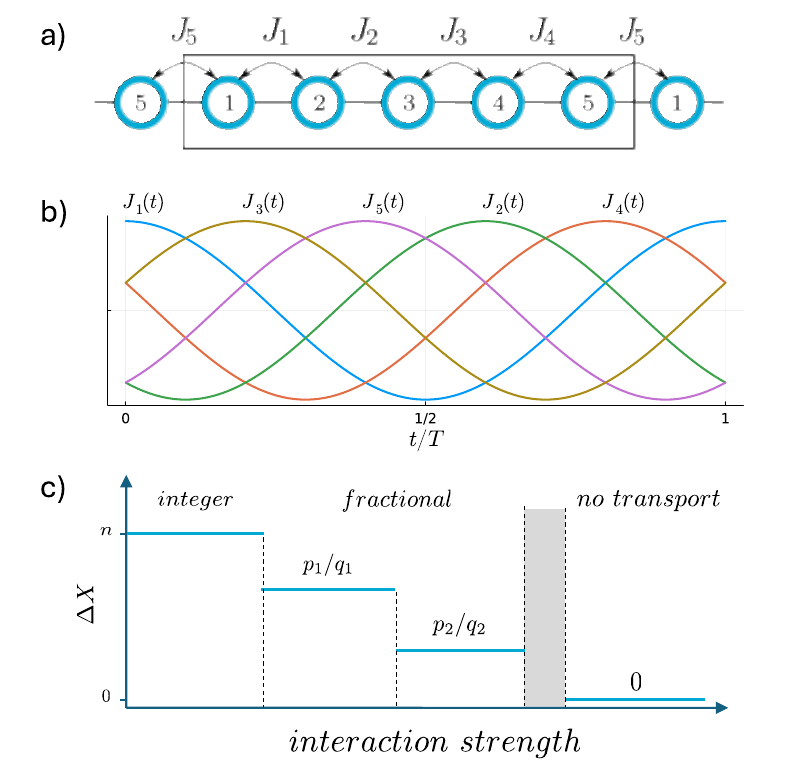}
    \end{centering}
    \caption{(a) and (b) 1D Aubry-Andr\'e-Harper model with modulated hopping rates $J_1(t) ... J_5(t)$ and on-site interactions $U$. (c) Motion of center of mass of a soliton $\Delta X$ 
  in a pump cylce. Upon increasing the interaction strength there are multiple transitions between phases of integer, fractional and eventually absent transport,
    in some cases intersected by a small interval of non-quantized transport (grey area). }
    \label{fig:system}
\end{figure}

Topological quantum systems have been intensively studied since the discovery of the quantum Hall effect  \cite{klitzing1980new}. 
One of the simplest examples for such systems is a Thouless pump  \cite{thouless_pump}, which displays a quantized  particle transport in an insulating bulk state of a 1D translationally invariant lattice upon cyclic adiabatic changes of system parameters. The transport is governed by an integer topological invariant, equivalent to a Chern number. 
The quantization of transport not only applies to a 
fully filled fermion band, it is also observable in the center-of-mass motion of a single particle equally distributed over all momentum states 
(see e.g. \cite{aidelsburger_review_thouless} for a detailed overview.) 
A major problem in the single particle case is the fast dispersion of the wavefunction. A possible solution for this has been utilized in \cite{lumer2013self,mukherjee2020observation,jurgensen2021quantized} using bound many-particle objects: lattice solitons. They show quantized transport in a topological pump  while being almost nondispersive due to their large mass. The notion solitons is used here colloquially as the self-bound many-particle states may not fulfill all properties of true solitons \cite{korepin1997quantum,mattis1986few,barbiero2014quantum}.
In the experiments of \cite{jurgensen2021quantized}, laser pulses have been injected into spatially modulated waveguide arrays simulating a time-dependent Aubry-Andr\'e-Harper (AAH) Hamiltonian 
with a Kerr nonlinearity mediating interactions \cite{aubry1980analyticity,harper1955single,kraus2012topological}. Increasing the light intensity, solitons form, for which integer transport in a full pump cycle was observed. Above a certain power threshold, all transport is halted. In subsequent work \cite{jurgensen2023quantized}, an interaction controlled  transition between phases with integer and \textit{fractional} transport was demonstrated (see Fig.\ref{fig:system}).

These observations triggered an extended body of theoretical work 
\cite{jurgensen2022chern,zhou2022topological,mostaan2022quantized,tuloup2023breakdown,hu2024pumping,sone2024nonlinearity,sone2025transition} based on the discrete nonlinear Schr\"odinger equation (DNLSE) \cite{lederer2008discrete}. While the DNLSE 
accurately reproduces the observed shift in the COM and relates the transition points between phases of different fractional transport to bifurcation points of soliton solutions, the underlying mean-field description fails to provide an explanation for the quantized transport in terms of a topological invariant and of the origin of the topological phase transitions. 

Using perturbation theory it has been argued in 
\cite{jurgensen2022chern,mostaan2022quantized}
that for weak interactions the position of the soliton follows that of the Wannier centers of the  single-particle band from which the soliton bifurcates, and thus is governed by the Chern number of this band.
However, as shown in Sect.~\ref{sec:mean-field}, there 
is a gradually growing admixture of higher single-particle bands 
when the interaction is increased, and the perturbative arguments fail entirely in the fractional case.
We note that there are approaches introducing nonlinear winding numbers \cite{zhou2022topological,sone2025transition} or Chern numbers \cite{sone2024nonlinearity} to characterize topological properties of the mean-field solutions also in the regime of large nonlinearities. These approaches are however limited to systems with certain symmetries or capture only special nonlinear solutions which do not include self-bound solutions such as solitons or time-dependent Hamiltonians.
Finally, cases of anomalous nonlinear Thouless pumping were found \cite{tao2024nonlinearity}, and
in \cite{tao2025nonlinearity} it was predicted  that multi-component solitons can show fractional transport, despite the fact that the single-particle bands are all topologically trivial. This shows that the topological transport of solitons cannot be traced back to topological properties of the underlying single-particle Hamiltonian.
Most recently a fermionic model with repulsive interactions has been investigated \cite{jurgensen2025fermions} where fractional transport emerges from coupling of single-particle bands by repulsive interactions of fermions.

We here unravel the origin of the quantized soliton transport and provide an explanation for the observed phase transitions by
developing a fully quantum description of topological soliton pumps.
To be specific, we consider the  AAH model as generic example; the approach applies, however, to all systems with self-bound many-particle states
in a translationally invariant lattice model.
We show that the topological contribution to their  transport  can be described in terms of an effective single-particle Chern number,
provided the soliton solution is gapped from all extended (not self-bound) many-body states. 
Introducing an effective single-particle Hamiltonian for solitons,
which due to translational invariance can be characterized by an effective soliton band structure, we show that in the AAH model for increasing interaction strength different soliton bands merge at some parameter values of the pump cycle. 
At this point, the transport is governed by a Wilson loop, giving rise to different phases with fractionally quantized average transport. Increasing the interaction further eventually mixes all soliton  bands, and since the total Wilson loop of all soliton COM  bands must vanish, the topological transport collapses
\cite{jurgensen2021quantized,fu2022nonlinear}. 
For single-particle Bloch Hamiltonians also particles prepared in excited bands show quantized topological transport, determined by the corresponding Chern number. 
We show that an analogous behavior can be observed for solitons if the system is prepared in an excited soliton band that is stable and has a finite band gap to lower (and higher) lying solutions.

Finally, we show that in some intermediate interaction regimes the energetically highest of a set of crossing soliton bands may be degenerate with bands of extended states. 
In such a case the transport is no longer (fractionally) quantized and may take arbitrary values, 
explaining the fluctuating transport numerically predicted in \cite{tuloup2023breakdown}.
Since the effective Hamiltonian of solitons is a single-particle Hamiltonian, its symmetries under time-reversal, charge-conjugation and chiral transformation provide a full classification of possible topological phases according to Ref.~\cite{altland1997nonstandard,ryu2010topological}.

\section{Model and mean-field approach} 

\subsection{Aubry-Andr\'e-Harper-model}

We consider a generic lattice model with attractive on-site interactions. Specifically, we investigate the bosonic tight-binding Hamiltonian
\begin{eqnarray}
    &&\mathcal{H}(t) = \mathcal{H}_0(t) + \mathcal{H}_\textrm{int}\label{eq:aah-hamiltonian}\\
    &&\enspace = -\sum_l \left [ \bigl(J_l(t) \hat{a}^{\dagger}_{l}\hat{a}_{l+1} + h.c.\bigr)+ \epsilon_l(t) \hat a_l^\dagger \hat a_l + \frac{U}{2} \hat{n}_{l}(\hat{n}_{l} - 1) \right ]\nonumber
\end{eqnarray}
where ${\cal H}_0$ describes the single-particle dynamics in the lattice and is periodic in time with period $T$, i.e. $\mathcal{H}_0(t)=\mathcal{H}_0(t+T)$. 
The corresponding hopping amplitudes $J_l$ and on-site energies $\epsilon_l$ have a spatial period $p$, which defines the unit cell size.
$\mathcal{H}_\textrm{int}$ describes an (attractive) on-site interaction of strength $U>0$, which will be parameterized
as $U=U_0/N$ with $N$ being the total number of particles. 
In order to understand the emergence of topological phase transitions observed in \cite{jurgensen2021quantized,jurgensen2023quantized} 
    we consider specifically the 1D Aubry-Andr\'e-Harper model \cite{aubry1980analyticity,harper1955single} with hopping amplitudes 
\begin{equation}
    J_l(t) = J\Bigl(1 + \delta \cos{\left(\Omega t + \frac{2\pi l k}{p}\right)}\Bigr),
\end{equation}
with $0<\delta < 1$,
 see Fig.~\ref{fig:system}.

The spatial period $p$ is chosen to be a prime number so that it is never commensurable with $l$ and/or $k$ which would effectively reduce the spatial period. The phase-offset $k$, which is chosen to be smaller than $p/2$, 
determines the smallest possible fraction in the topological transport.
If not explicitly stated otherwise, we choose a spatial period $p=5$ and a phase-offset $k=2$. This is the smallest possible unit-cell size where fractional transport appears in the AAH-chain.
All on-site potentials $\epsilon_l$ in Eq. \eqref{eq:aah-hamiltonian} are chosen to be $0$. 

With periodic boundary conditions, the Hamiltonian is 
translational invariant, i.e.
$\hat T\mathcal{H} \hat T^{-1} = \mathcal{H}$, where $\hat T$ is the translation operator by one unit cell, i.e.
$\hat T \hat a_l \hat T^{-1} = \hat a_{l+p}$. 
As a consequence, the lattice momentum $K$ of the center of mass (COM) is a conserved quantity.

Attractive interactions $U$ lead to localized soliton states.  These are states with a distribution of occupation numbers that decay with increasing distance to the center of mass
with a localization length $\xi$, i.e. 
$    \langle \hat n_{l+d}\hat n_{l}\rangle \, \sim \, \exp\{-\vert d\vert /\xi\}$ for $d \gg 1.$
We call them \textit{stable} if they have an energy gap to all extended states with the same $K$. 
On the quantum level, a minimum value $U_c$ is required for a soliton to form, which tends to zero as $N\to\infty$ \cite{naldesi2019rise}. In a complex energy structure multiple soliton solutions can exist, which for some parameter values may become degenerate.
Furthermore, stable excited soliton solutions may not exist at all times, as they can become degenerate with extended states in some parts of the pump cycle.
 In this case we call these solitons partially stable.

  \subsection{Mean-field approach: Discrete nonlinear Schr\"odinger equation}
 \label{sec:mean-field}

In mean-field approximation the dynamics of solitons in a 1D lattice are described by the discrete nonlinear Schr\"odinger equation. This equation  can be obtained from a Gutzwiller coherent-state ansatz \cite{gutzwiller} for the many-body quantum state
\begin{equation*}
    \vert \psi\rangle = \prod_l \vert \phi_l\rangle,\qquad \textrm{where}\qquad\hat a_l\vert \phi_l\rangle = \phi_l \vert \phi_l\rangle. 
\end{equation*}
For the AAH Hamiltonian, Eq. \eqref{eq:aah-hamiltonian}, it reads
($\hbar =1$)
\begin{eqnarray}
    i\frac{\partial}{\partial t} \phi_l = - J_l \phi_{l+1} - J_{l-1} \phi_{l-1} - \epsilon_l \phi_l - U \vert \phi_l\vert^2 \phi_l.\label{eq:DNLSE}
\end{eqnarray}
Numerical simulations of Eq.~\eqref{eq:DNLSE} have provided a good description of soliton energies and the observed soliton transport in parameter regimes where the soliton is stable. The semiclassical description, however, fails to explain the topological nature of the transport and the origin of topological phase transitions.
In \cite{jurgensen2022chern,mostaan2022quantized} it was argued that 
for weak interactions the solitons can be represented in terms of the most localized Wannier states of 
single-particle Bloch bands neglecting exchange terms between the latter. 
Under this assumption the topological transport is described by the Chern number of the Bloch band the soliton bifurcates from.

\begin{figure}[H]
    \centering
    \includegraphics[width=0.48\textwidth]{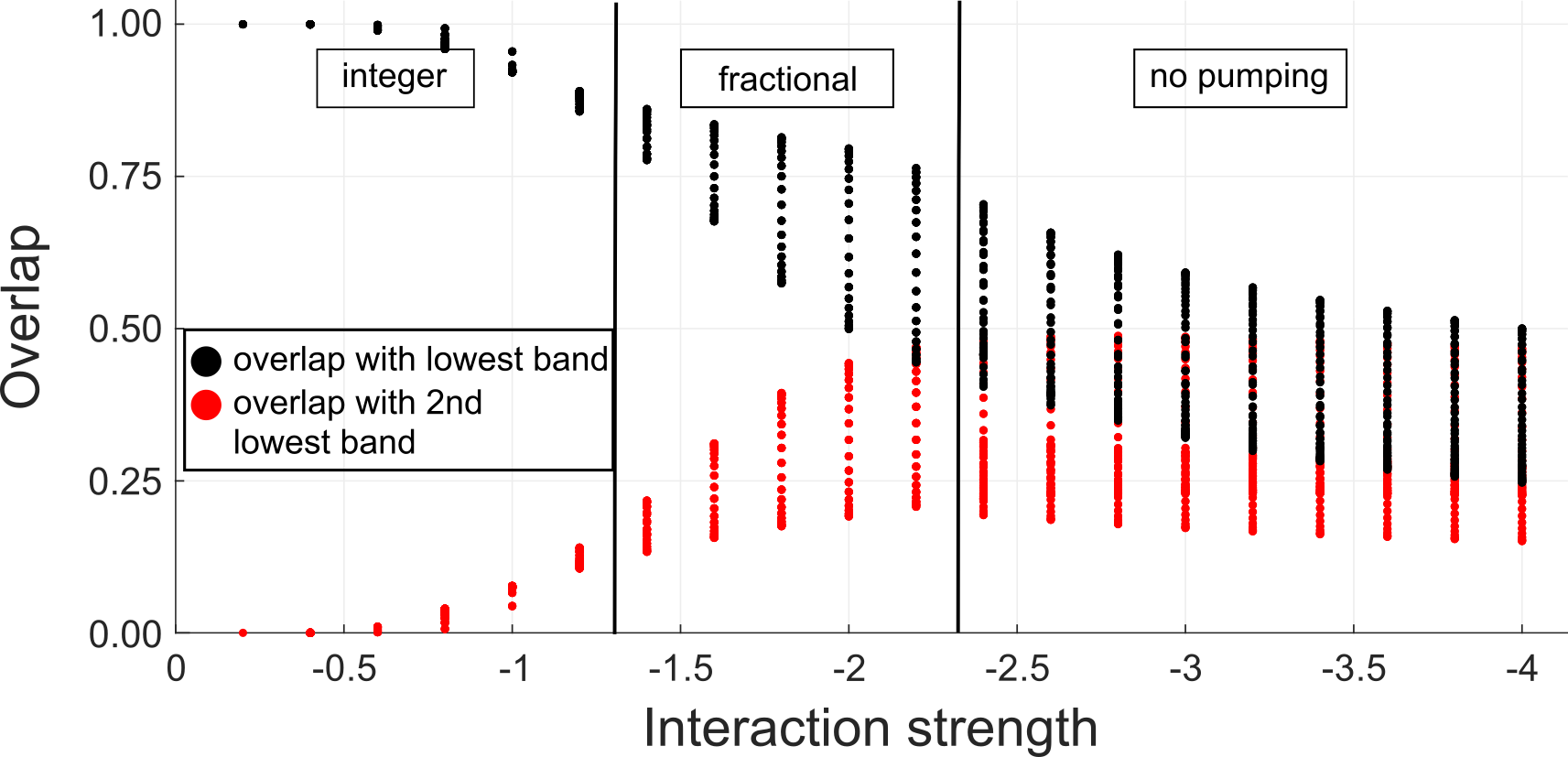}
    \caption{Overlap of eigensolution of the DNLSE, Eq. ~\eqref{eq:DNLSE}, for the AAH model with phase offset $k=2$ and unit-cell size $p=5$ with single-particle Bloch bands for different interaction strength $U/J$. Data points for the same interaction strength correspond to different times in the pumping cycle.}
    \label{fig:overlap}
\end{figure}

In Fig.~\ref{fig:overlap} we have plotted the integrated overlap of a soliton  obtained from a self-consistent solution of Eq.~\eqref{eq:DNLSE} with the single particle Bloch wavefunctions for the two lowest bands for the AAH model with phase offset $k=2$ and unit cell size $p=5$ as a function of interaction strength $U/J$. 
The vertical spread of data points reflects different times in the pump cycle. One recognizes that while in the perturbative limit of very small absolute values of interaction strengths there is indeed a 
close to unity overlap with the lowest Bloch band, contributions from higher bands continuously grow with increasing interaction strength. 
In particular, both in experiments and in numerical simulations of the DNLSE, integer transport is seen up to rather large interaction strength at which the admixture of higher Bloch bands is already quite substantial. Moreover,
in the regime of fractional transport, following the above argument, one would expect equal contributions from the lowest and the first excited band, which is clearly not the case. Therefore, the resulting topological invariant is not just the average of the single-particle Chern-numbers.

\section{Center-of-mass transport and many-body Chern number}

Since the semi-classical approaches to describe the transport of lattice solitons were shown to be incomplete, we will here derive a quantum mechanical description. Specifically, we are interested in the transport of the COM of a soliton when 
${\cal H}_0(t)$ is adiabatically varied over a period $T$, where the translational invariance is not changed. The instantaneous eigenstates of ${\cal H}(t)$ can be classified in terms of the conserved COM momentum $K$ and a band index $\mu$, and will be denoted as $\vert E_\mu(K)\rangle$ with energy $E_\mu(K)$.

The time evolution of the center of mass position of the $N$ particles, $\hat X =\sum_{j=1}^N \hat x_j/N$, is governed by the $N$-particle velocity operator $\partial_t \hat X = \hat V = -i[\hat X,{\cal H}] = -i [\hat X,{\cal H}_0]$, which can be conveniently expressed in terms of a momentum-shifted Hamiltonian $\mathcal{H}(q) = e^{-i q \hat X}\, \mathcal{H}\,  e^{i q\hat X}$. I.e. $\hat V = \hat V(q)\vert_{q=0}$ with $\hat V(q) = e^{-i q \hat X}\, \hat V\,  e^{i q\hat X} = \partial {\cal H}(q)/\partial q$. 
\begin{equation}
    \frac{d}{dt} \langle \hat X\rangle = \left\langle \hat V(q)\right\rangle_{q=0}=\left \langle \frac{\partial \mathcal{H}(q)}{\partial q}\right\rangle_{q=0}.
\end{equation}
In the following we discuss the topological contributions to $d\langle\hat X\rangle/dt$ for degenerate and non-degenerate solutions separately.

\subsection{Non-degenerate soliton solutions}

 Let us first consider a non-degenerate soliton solution, i.e. for a given COM momentum $K$ there is only one eigenstate for each energy value in the whole pump cycle. In order to calculate the transport of a soliton from a single non-degenerate band, one can follow the well-known arguments for the transport of a single particle \cite{thouless_pump,thouless1982quantized}. To account for the topological transport starting in an instantaneous eigenstate  $\vert E_0(K,t)\rangle$ at time $t$, one has to take into account non-adiabatic corrections. As the modulation of the Hamiltonian does not affect the translational invariance, non-adiabatic transitions only couple to states with the same COM  momentum. Non-degenerate time-dependent perturbation theory yields
 \begin{equation}
     \vert \Psi(K,t)\rangle = \vert E_0(K) \rangle  + i \sum_{\alpha\ne 0}
    \vert E_{\alpha}(K)\rangle\frac{\langle E_{\alpha}(K)\vert \partial_t E_0(K)\rangle }{E_{\alpha}(K)-E_{0}(K)}
 \end{equation}
where we have suppressed the dependence on time for notational simplicity. We note that $\vert \partial_t E_0(K)\rangle $ is orthogonal to $\vert E_0(K)\rangle $ and that the second term in the above expression is small. Calculating the average velocity in this states gives in lowest order
of perturbation theory
\begin{eqnarray}
    &&\langle \hat V(q=0,t)\rangle = 
    \frac{\partial E_{0}(K)}{\partial K} \nonumber\\
    && \qquad + i \sum_{\alpha\ne 0}  
    \Biggl(\frac{\langle E_{0}\vert \partial_q \mathcal{H}(q)\vert_{q=0}\vert E_{\alpha}\rangle\langle E_{\alpha}\vert \partial_t E_{0}\rangle }{E_{\alpha}-E_{0}} -c.c.\Biggr)\nonumber\\
    && \qquad=\frac{\partial E_{0}(K)}{\partial K} + i \left(\left\langle \frac{\partial E_0}{\partial t }\biggr\vert \frac{\partial E_0}{\partial K}\right\rangle- c.c. \right).\label{eq:v}
\end{eqnarray}
For the second step we used
\begin{eqnarray}
    \label{eq:identity1}
    0 && =  \partial_q\langle E_{0}\vert \mathcal{H}(q)\vert E_{\alpha}\rangle \\ 
    && = \langle E_{0}\vert \partial_q \mathcal{H}(q)\vert E_{\alpha}\rangle + E_0 \langle E_{0}\vert \partial_q E_{\alpha}\rangle + E_{\alpha} \langle \partial_q E_{0}\vert E_{\alpha}\rangle \nonumber
\end{eqnarray}

\noindent as well as
\begin{eqnarray}
    0 = \partial_q\langle E_{0}\vert E_{\alpha}\rangle = \langle E_{0}\vert \partial_q E_{\alpha}\rangle + \langle \partial_q E_{0}\vert E_{\alpha}\rangle.
    \label{eq:identity2}
\end{eqnarray}

The first term in expression Eq.~\eqref{eq:v} is just the group velocity of the band and  describes the dynamical contribution to the transport. If we consider an initial  
soliton state with coefficient $c(K)$ in momentum state $\vert E_0(K)\rangle$, which is symmetric with respect to $K=0$, the dynamical contribution vanishes. 
Then, the shift of the center of mass of the soliton in one period $T$ is given by the second term in Eq.~\eqref{eq:v} integrated over all COM momenta $K$
\begin{equation}
    \Delta \langle \hat X\rangle = 
     \int_0^T\!\!\! dt 
    \int_{-\pi}^{\pi}\! \!\! dK\,  \vert c(K)\vert^2
    \, {\cal F}(K,t).\label{eq:transport-general}
\end{equation}
Eq.~\eqref{eq:transport-general} is a weighted integral over the Berry-curvature 
\begin{equation}
    {\cal F}(K,t) = i\left(\left\langle \frac{\partial E_0}{\partial t }\biggr\vert \frac{\partial E_0}{\partial K}\right\rangle- c.c. \right)
\end{equation}
in $(K,t)$ space.
In the special case, where the COM wavefunction of the soliton is initially localized to a single unit cell, all coefficients are equal, $\vert c(K)\vert^2 = 1/{2\pi}$. In this case the transport is given by the effective single-particle Chern number of the soliton Bloch band, i.e. it is integer quantized:
\begin{equation}
    \Delta \langle \hat X\rangle = 
     \int_0^T\!\!\! dt 
    \int_{-\pi}^{\pi}\! \frac{dK}{2\pi}\, {\cal F}(K,t) =C.
    \label{eq:COM-Chern-number}
\end{equation}

Note that we did not make any assumption about the many-body wavefunction other than its gapfulness. In particular, no assumption about its behavior in the \textit{relative} coordinates of the particles was made. However, the assumption of gapfulness of the $N$-particle state with fixed COM momentum $K$ requires in general that the $N$ particles are bound to each other.\\
Finally we comment on the relation between our description of soliton transport and other topological invariants for interacting many-body systems, such as the Niu-Thouless (NT) invariant \cite{niu_chern_number}. The NT invariant employs twisted boundaries and is an integral over a Berry curvature in $(\theta,t)$ space, with $\theta\in \{0,2\pi\}$ being the twist angle. In contrast to ${\cal F}(K,t)$, see Eq.~\eqref{eq:transport-general}, a Berry curvature in $(\theta,t)$ would not describe the transport of an individual soliton but applies only to an insulating many-body state, such as a band insulator of non-interacting fermions, where all momentum states are equally occupied.

\subsection{Degenerate soliton solutions}
\label{sec:degenerate-wilson-loop}

 In the degenerate case, i.e. if there are crossings of $m$ solutions at some points in time, 
we need to apply degenerate time-dependent perturbation theory 
to an $m$ component vector $\vert \underline{\Psi}(K)\rangle =\bigl( \vert \Psi_0(K)\rangle,\dots, \vert \Psi_{m-1}(K)\bigr)^\top$ and can express the time-evolved state 
as
\cite{wilczek-zee}
\begin{eqnarray}
    \vert \underline{\Psi}(K,t)\rangle &=& {\cal T}\exp\left\{-i\int_0^t\!\! d\tau \mathbf{A}_K(\tau)\right\} \vert \underline{\Psi}(K,0)\rangle\qquad \label{eq:Wilczek-Zee} \\
    && + \textrm{non-adiabatic\, terms},\nonumber
\end{eqnarray}
where
\begin{eqnarray}
    &&\mathbf{A}_K(\tau) = \nonumber \\
    &&\left(
    \begin{array}{cccc}
         E_0 &  i \langle E_0\vert \partial_t E_1\rangle &  i \langle E_0\vert \partial_t E_2\rangle & ... \\
          i \langle E_1\vert \partial_t E_0\rangle & E_1 &  i \langle E_1\vert \partial_t E_2\rangle & ... \\
          i \langle E_2\vert \partial_t E_0\rangle &  i \langle E_2\vert \partial_t E_1\rangle & E_2 & ... \\
         ... & ... & ... & ...
    \end{array}
    \right)
    \label{eq:wilczek-zee-matrix}
\end{eqnarray}
is the Wilczek-Zee non-Abelian Berry connection, and ''non-adiabatic terms'' denotes the perturbative contributions due to non-adiabatic couplings to other, energetically separated states similar to the non-degenerate case. We note that since
$\langle E_l\vert \partial_t E_m\rangle = \bigl(\langle E_l\vert \partial_t {\cal H} \vert E_m\rangle\bigr)/(E_m-E_l)$ (compare Eq. \eqref{eq:identity1} and \eqref{eq:identity2}), the eigenstates of the matrix $\mathbf{A}_K$ coincide  with the bare states $\vert E_0\rangle \dots \vert E_m\rangle$ far away from the crossing point in the adiabatic limit. At the crossing point the off-diagonal elements however diverge in general, which leads to a mixing. 

Suppose the cyclic change of the Hamiltonian starts at a point where there is no degeneracy between COM bands and the system is prepared in one solution, say $\vert E_0(K)\rangle$. Then, in the presence of isolated crossing points with other solutions 
$\vert E_\alpha(K)\rangle$, a single cycle $t=0 \to t=T$ will in general not return the initial state to itself, but multiple cycles are needed. 
Therefore, the topological transport is only integer quantized after multiple cycles, giving rise to a fractional average transport per cycle.
In this case where there are crossings of soliton solutions at some point in time, say of $\vert E_0(K,t)\rangle$ and $\vert E_1(K,t)\rangle$, the Chern number must be generalized to a Wilson loop
\begin{equation}
   {C}_n= \frac{1}{2\pi} \int_0^{T} dt \, \partial_t \text{Im log det } \mathbf{W}(t)
    \label{eq:wilson-loop}
\end{equation}
where $\mathbf{W}(t) = {\cal T}\exp\left\{i\int_\textrm{BZ} dK \,\mathbf{B}_t(K)\right\}$, and 
\begin{eqnarray}
    \mathbf{B}_t = \left(
    \begin{array}{cc}
         \langle E_0 \vert \partial_K E_0\rangle &  i \langle E_0\vert \partial_K E_1\rangle \\
          i \langle E_1\vert \partial_K E_0\rangle & \langle E_1 \vert \partial_K E_1\rangle
    \end{array}
    \right)\label{eq:Wilczek-Zee-1}
\end{eqnarray}
is the Wilczek-Zee non-Abelian Berry connection \cite{wilczek-zee} for fixed time, here for $n=2$ crossing bands.
$C_n$ is an integer and the \textit{average} topological transport per cycle in the $n$ bands is given by $C_n/n$.

The COM transport of a soliton is then \textit{fractional}, provided it returns to its original energy only after $n$ periods. We will show that this is the case for the Aubry-Andr\'e-Harper model of Ref. \cite{jurgensen2023quantized}.

\section{Calculating Chern number and Wilson loop of solitons}

While the soliton energy structure can be well approximated by a self-consistent solution of the DNLSE, the many-body eigenstates and the Chern number, Eq. \eqref{eq:COM-Chern-number}, or the Wilson loop, Eq. \eqref{eq:wilson-loop}, must be obtained from solving the many-body Schr\"odinger equation, which constitutes a substantial challenge for more than a few particles. To tackle this problem 
we introduce the following basis of states with a fixed $K$ \cite{scott1994quantum,ke2017multiparticle} 
\begin{equation}
    \vert \Psi_\alpha(K)\rangle= \sum_{m=0}^{L-1} \bigl(e^{i K} \hat T\bigr)^m 
    \, \vert \Phi_\alpha(0;K)\rangle.
\end{equation}
Here we assume a lattice with $L$ unit cells, each containing $p$ sites, and periodic boundary conditions. The states
\begin{equation*}
    \vert \Phi_\alpha(0;K)\rangle = 
    {\sum_{\{n_l\}}}^\prime\, c_\alpha\bigl[\{n_l\};K\bigr]\, \, 
    \bigl\vert 
    \{n_l\}
    \bigr\rangle
    \label{eq:relative_distribution_state}
\end{equation*}
describe the distribution of particles $n_l = \{n_{-pL/2 + 1}\dots n_{pL/2}\}$ around the lattice site $l=0$ (conveniently chosen close to the center of mass of all particles), with coefficients $c_\alpha[\{n_l\},K]=c_\alpha[n_{-pL/2 + 1}\dots n_{pL/2};K]$ and $\bigl\vert\{n_l\}\bigr\rangle$ being a number state. We assume for simplicity that $L$ is even. $\sum^\prime_{\{n_l\}}$ denotes summation over all $n_l$'s for which $\sum_l n_l = N$. Translation by a unit cell gives $\hat T\,  \vert \Phi_\alpha(0,K)\rangle = \vert \Phi_\alpha(p,K)\rangle$.
(Note that in order to guarantee orthonormality, states $\vert n_{-\frac{pL}{2}+1},.., n_{\frac{pL}{2}}\rangle $ must 
not be eigenstates of $\hat T$.)
Now we can make use of the fact that for large attractive interactions the solitons have a small localization length $\xi$. Thus we can restrict ourselves to special cases: (i) two-site solitons where the basis states have at most two (neighboring) sites populated (i.e. $n_0, n_1 = N - n_0 \neq N $ in general) and (ii) three-site solitons where three adjacent lattice sites might be populated (i.e. $n_0, n_1, n_2 = N-(n_0+n_1)\neq N $ in general).
In the basis of COM-momenta $K$ the many-body Hamiltonian (Eq. \eqref{eq:aah-hamiltonian}) is block diagonal, i.e. the COM-momenta are decoupled. For the two-site solitons the block dimension is $p \cdot N$, for the three-site soliton 
it is $p \cdot N (N+1)/2$, which allows us to perform numerical simulations for tens to hundreds of particles. 
Due to the limitation to two-site or three-site solitons, some of the higher energy solutions are 
not true eigensolutions, but are enforced by the boundary conditions. These states can be detected, however, by comparing two and three-site solutions
and are not relevant in the low-energy regime, discussed here.

\begin{figure}[H]
    {\centering
    \begin{subfigure}[t]{0.235\textwidth}
        \includegraphics[width=\textwidth]{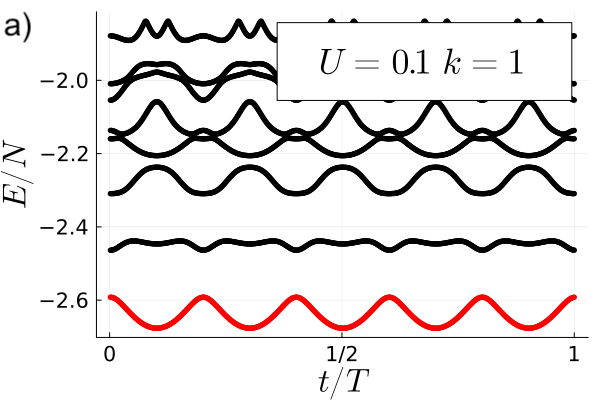}
    \end{subfigure}
    \begin{subfigure}[t]{0.235\textwidth}
        \includegraphics[width=\textwidth]{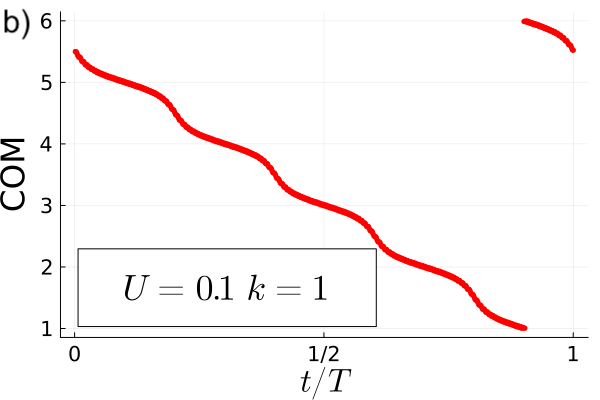}
    \end{subfigure}
    \begin{subfigure}[t]{0.235\textwidth}
        \includegraphics[width=\textwidth]{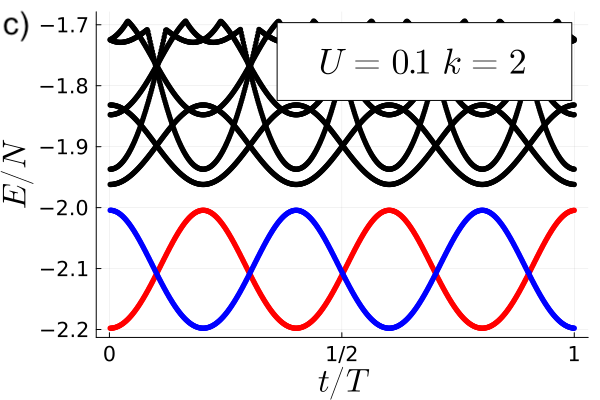}
    \end{subfigure}
    \begin{subfigure}[t]{0.235\textwidth}
        \includegraphics[width=\textwidth]{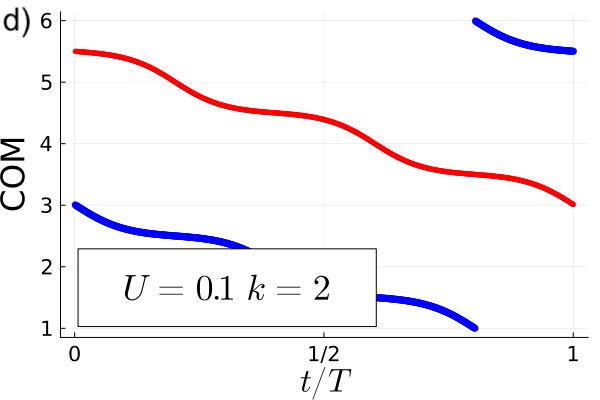}
    \end{subfigure}
    }
    \caption{Instantaneous soliton energies for  10 particles for attractive interaction $U/J=0.1$ with phase-offset $k=1$ in (a) and with phase-offset $k=2$ in (c),  and unit-cell size $p=5$. Note that the soliton bands are almost flat and the variation of $E(K)$ with $K$ is less than the width of the lines. The corresponding COM-movements in a pump cycle of the red and blue marked energies are shown in (b) and (d). The energies and COM-positions are obtained with the three-site soliton ansatz.} 
    \label{fig:hugo-phases}
\end{figure}

In Fig.~\ref{fig:hugo-phases} we have plotted the energies of the \mbox{solitons}  for the AAH model, Eq.~\eqref{eq:aah-hamiltonian}, and the COM transport as function of time, for $N=10$ particles with unit cell size $p=5$ and $U=0.1$ for phase offsets $k=1$ and $k=2$. 
(Note that the soliton bands are almost flat with a bandwidth less than the width of the lines.)
In the first case (Fig.~\ref{fig:hugo-phases}~a,~b) there is a single lowest-energy soliton solution with many-body Chern number, Eq. \eqref{eq:COM-Chern-number}, $C=1$ and integer transport. In the second case (Fig.~\ref{fig:hugo-phases}~c,~d) two energies cross with 
a non-trivial Wilson loop, Eq. \eqref{eq:wilson-loop}, $C_2=1$ giving rise to fractional transport of $1/2$. The data is calculated within the three-site soliton ansatz.

\begin{figure}[H]
    {\centering
    \includegraphics[width=0.5\textwidth]{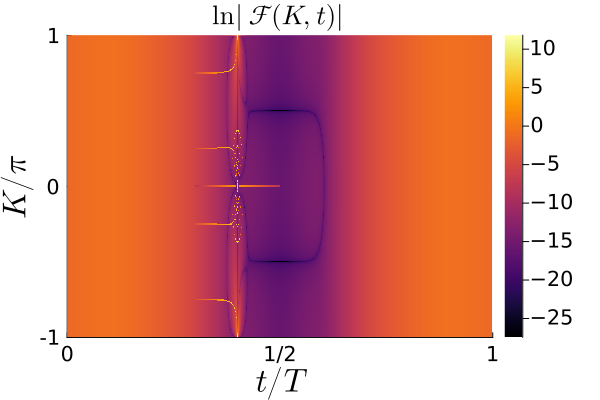}}
    \caption{Logarithm of the absolute value of the Berry curvature $\ln|{\cal F}(K,t)|$ in the lowest energy solution for $N=3$ particles, a $p=5$ unit-cell and a $k=1$ phase-offset. Integrating Eq.~\ref{eq:COM-Chern-number} with this Berry curvature yields a Chern number $C=1$. The sharp peaks, where $\ln|{\cal F}(K,t)|$ is on the order of 10, are numerical relicts but are effectively a set of measure zero for the integration.}
    \label{fig:berry-curvature}
\end{figure}

With the same approach we can numerically calculate the Berry curvature ${\cal F}(K,t)$ and from this the Chern number or Wilson loop. In Fig.~\ref{fig:berry-curvature} we have given an example for the Berry curvature of the AAH-model with $N=3$ particles, a $p=5$ unit-cell and a $k=1$ phase-offset, which shows integer quantized transport similar to Fig. \ref{fig:hugo-phases}. Using Eq.~\ref{eq:COM-Chern-number} for this Berry curvature yields a COM Chern number $C=1$. The yellow lines in Fig. \ref{fig:berry-curvature} are numerical relicts and do not influence the result of the integration since they are effectively sets of measure zero.

We here showed the applicability of our COM approach to identify the different topological phases. In the following we will provide an explanation for the phase transitions.

%
\section{Topological phase transitions}
\label{sec:phase_transitions}

In the following we provide an explanation for the  transitions between phases with integer, fractional and vanishing transport observed in the experiments.

\subsection{Effective soliton Hamiltonian}
\label{sec:effective_hamiltonian}

 The COM dynamics of bound $N$-particle objects (solitons) can be described by an effective single-particle Hamiltonian with eigenstates $\vert E_\mu(K)\rangle$:
\begin{equation}
    {\cal H}_\textrm{eff} = \sum_\mu \sum_K  E_\mu(K) \, \vert E_\mu(K)\rangle\langle E_\mu(K)\vert.
\end{equation}
Since such a model misses out the continuum of extended states, it is only adequate for stable soliton solutions. Defining annihilation and creation operators for solitons centered at lattice site $l$ as $\hat d_l$ and $\hat d_l^\dagger$, respectively, the effective soliton Hamiltonian would read in coordinate space
\begin{equation}
    \mathcal{H}_{\textrm{eff}}(t) = - \sum_l \left [ \bigl(J_{l, \textrm{eff}}(t)\,  \hat{d}^{\dagger}_{l}\hat{d}_{l+1} + h.c.\bigr) + \epsilon_{l, \textrm{eff}}(t) \, \hat{d}^{\dagger}_{l}\hat{d}_{l} \right ].
    \label{eq:effective_hamiltonian}
\end{equation}
where the $J_{l, \textrm{eff}}$ are the effective hopping rates of the soliton and the $\epsilon_{l, \textrm{eff}}$ effective local energies. (Note that we here have assumed only nearest neighbor hopping.)
All topological properties, including the topological classification according to the Altland-Zirnbauer scheme \cite{altland1997nonstandard,ryu2010topological}, as well as all phase transitions of solitons are determined by this effective Hamilton.

In the limit of strong attractive interactions all soliton solutions have a localization length of a single lattice constant, i.e. all particles are located at the same lattice site. In this limit, we can explicitly derive the effective Hamiltonian from perturbation theory. Furthermore, it is obvious that in this limit  the number of soliton bands is the same as the number of single-particle bands. 
We will explicitly construct $\mathcal{H}_{\textrm{eff}}$ for $N=3$ particles in the strong interaction limit in Sec.~\ref{sec:triplon}. 
Here we will first discuss some of its general properties.

In the Aubry-Andre-Harper model, Eq. \eqref{eq:aah-hamiltonian}, any small attraction $U_0$ is sufficient to form a bound state (lattice soliton) with an energy  below the continuum of extended $N$-particle states, if $N$ is large. Increasing the attractive interaction, the energy of these solitons is lowered and the bands deform. Moreover, excited soliton solutions can emerge. If the soliton bands have a non-trival Chern number or Wilson loop, (fractional) quantized topological transport can be observed
\cite{jurgensen2021quantized,jurgensen2023quantized}.

\subsection{Merging of soliton bands}

The experiments in \cite{jurgensen2021quantized,jurgensen2023quantized} and DNLSE simulations showed  that for very large but still finite values of the interaction, the topological transport stops altogether. This can be understood as follows: For $U\gg J$ the localization of all soliton solutions is reduced to a single lattice site. Thus the contribution of the local interaction to the 
energy, $U N(N-1)/2$, is the same for all solitons . 
Using a perturbative ansatz, one can show that moreover the contribution of the kinetic energy can be disregarded as it scales as $\sim J (\frac{J}{U})^{N-1}$ (see Appendix and Ref.~\cite{Mila_2010}). Thus, the effective soliton Hamiltonian becomes approximately diagonal and is entirely determined by local energies $\epsilon_{l,\textrm{eff}}$, which for the AAH model result from virtual hopping processes of a single particle from the soliton site $l$ to an empty neighboring site (and back) and are in second-order perturbation theory  proportional to $\epsilon_l\sim (J_l^2(t) + J_{l+1}^2(t))/U$.
As a consequence, all soliton bands cross at different points of the pump cycle and the topological transport is given by the total Wilson loop of all COM bands, which is always zero for general reasons.

%
\begin{figure}[H]
    \begin{center}
        \includegraphics[width=0.4\textwidth]{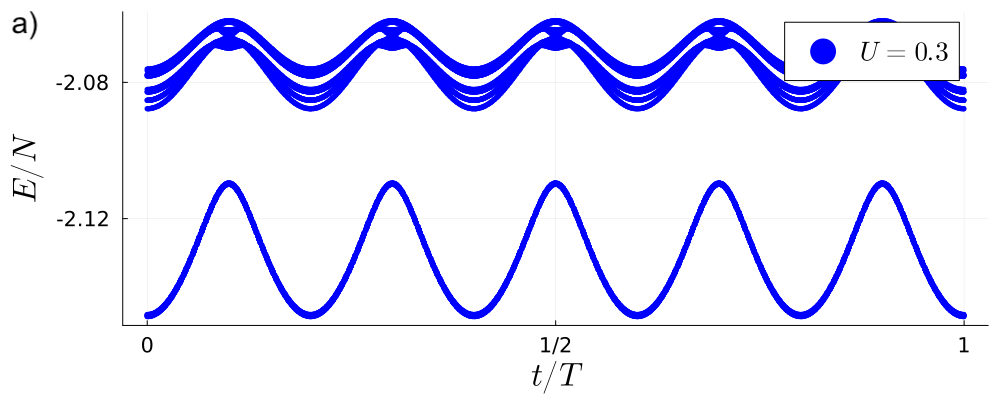}
    \includegraphics[width=0.4\textwidth]{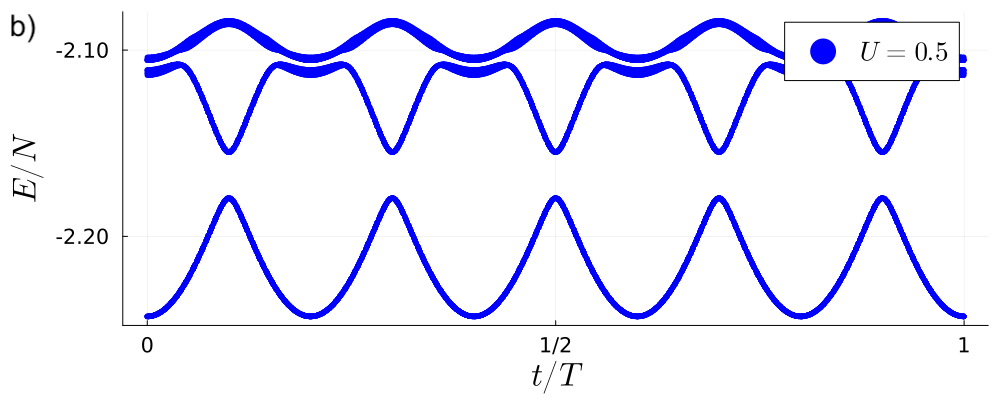}
    \includegraphics[width=0.4\textwidth]{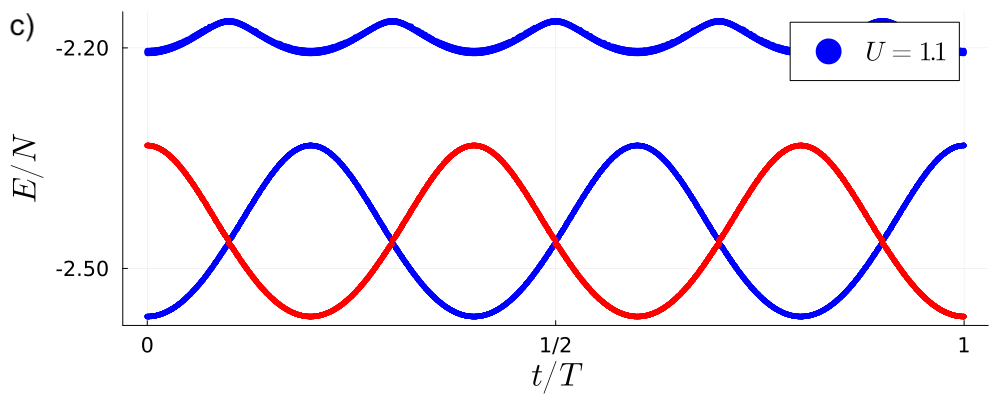}
    \end{center}
    \caption{Merging of soliton energies obtained from exact diagonalization with increasing interaction strength $U$ for $N=3$ particles, a $p=5$ unit-cell and $k=2$ phase-offset. The number of unit cells is $L=6$. 
    A soliton originally prepared at $t=0$ in one of the two bands will remain in this band if the changes are adiabatic and the soliton energies do not touch (Figs. a) and b)). Once the energies touch in Dirac-like cones, the solutions switch bands at every crossing (red curve in Fig. c). The width of the lines is larger than the width of the soliton bands in $K$ space.}
    \label{fig:merging-bands}
\end{figure}
%

The two limiting cases show that as $U$ is increased, initially separated soliton solutions must approach each other and  eventually merge. As shown in Fig.~\ref{fig:merging-bands}, obtained from exact diagonalization (ED) simulations of a small system (30 lattices sites), they do so by forming Dirac-like cones which touch 
at a critical interaction $U_{c}$. In this case the Chern numbers of the lowest two soliton bands are no longer adequate topological invariants, but one has to consider the Wilson loop. The topological transport can then become fractional, provided the soliton solutions which started at $t=0$ are energetically well separated  and switch bands at the Dirac-like points. We now show that this is indeed the case:

When the two soliton bands touch, the adiabatic evolution is governed by 
a $U(2)$ transformation \cite{wilczek-zee} (compare Eq. \eqref{eq:Wilczek-Zee} in Sec.~\ref{sec:degenerate-wilson-loop})
\begin{eqnarray}
    \vert \underline{\Psi}(K,t)\rangle &=& {\cal T}\exp\left\{-i\int_0^t\!\! d\tau \mathbf{A}_K(K,\tau)\right\} \vert \underline{\Psi}(K,0)\rangle. \,\,\, 
\end{eqnarray}
Close to the crossing points, which we assume to take place at $t=t_0$, the non-Abelian Berry connection $\mathbf{A}_K$ takes on the form 
\begin{equation}
      \mathbf{A}_K = \left(
    \begin{array}{cc} a\vert \tau\vert 
          & - \frac{i b}{\tau}  \\
           \frac{i b}{\tau}   & - a \vert \tau\vert
    \end{array}
    \right),
\end{equation}
with $\tau = t-t_0$. Here,
we assumed that apart from a common energy offset  $E_0 =-E_1= a\, \vert \tau\vert$, and thus
$\langle E_0\vert \partial_t {\cal H} \vert E_1\rangle /(E_1-E_0)\approx b / \tau$. 
Due to the diverging off-diagonal elements, the eigensolutions no longer follow the original solutions but cross from one  to the other (see color code in Fig.~\ref{fig:merging-bands}~c). Since $p$ is a prime number, there is an odd number of crossing points in one period in the AAH model. Thus, two periods are required for a soliton to return to the solution it started from. The shift in the COM position is then integer quantized only after two periods and the average transport per cycle is fractionally quantized. 
This integer-to-fractional transition is sharp, i.e. it does happen without a region of non-quantized transport in-between the two phases. The following section will however discuss a situation where an intermediate region of non-quantized transport appears.

\subsection{Failure of transport quantization}

\begin{figure}[H]
    \centering
    \begin{subfigure}[c]{0.03\textwidth}
        \includegraphics[width=\textwidth]{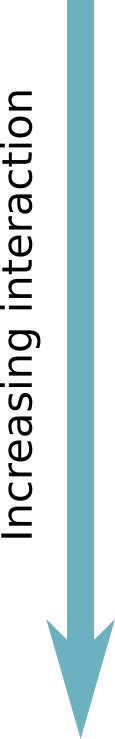}
    \end{subfigure}
    \begin{subfigure}[c]{0.4\textwidth}
        \begin{subfigure}[t]{\textwidth}
            \includegraphics[width=\textwidth]{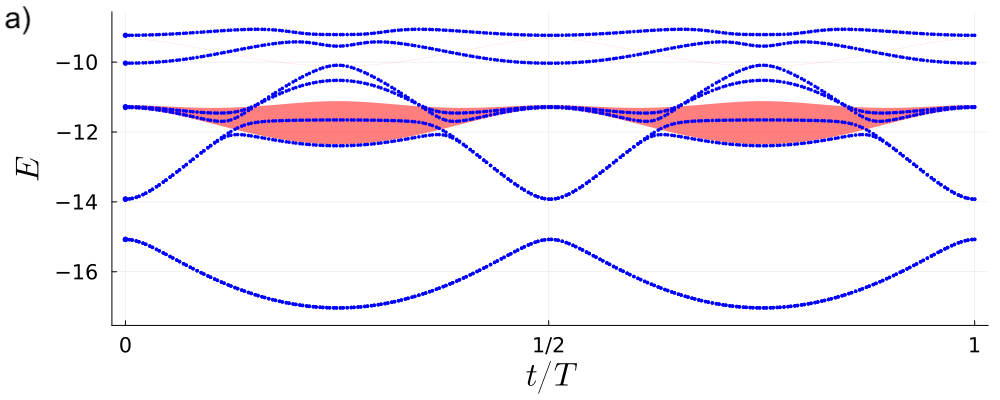}
        \end{subfigure}
        \begin{subfigure}[t]{\textwidth}
            \includegraphics[width=\textwidth]{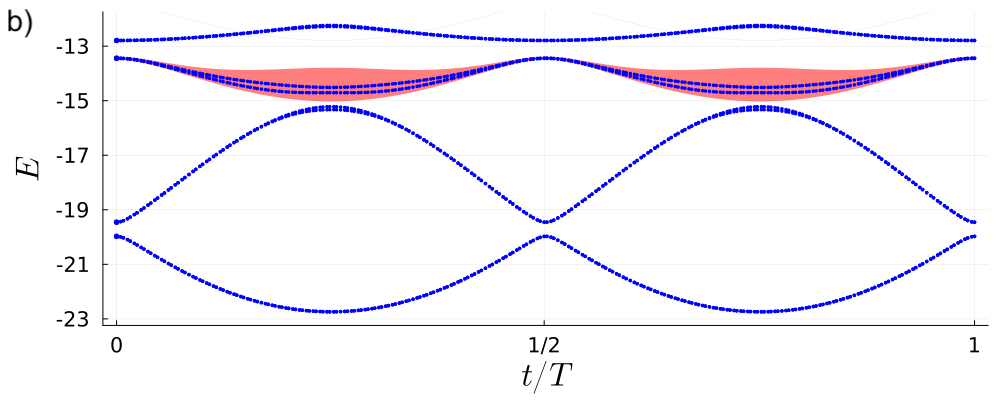}
        \end{subfigure}
        \begin{subfigure}[t]{\textwidth}
            \includegraphics[width=\textwidth]{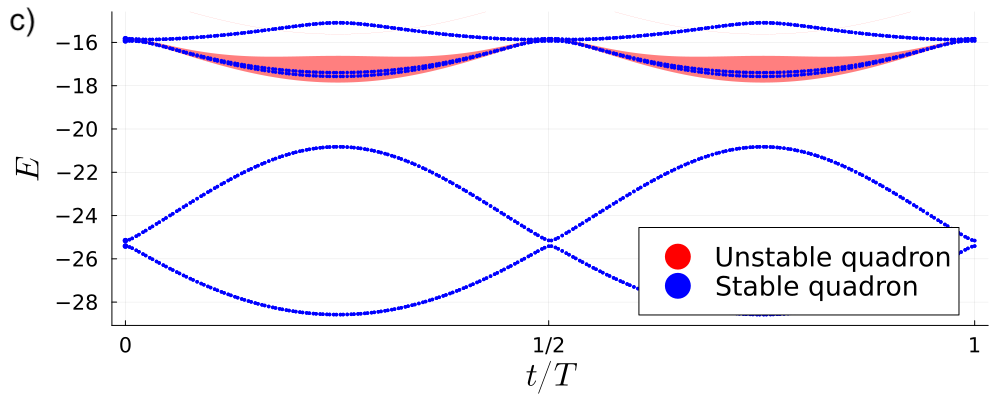}
        \end{subfigure}
    \end{subfigure}
    \caption{Breakdown of transport quantization in the Rice-Mele model for intermediate values of onsite interactions. Shown are lowest energies for $N=4$ bosons. a) $U=2.0$: lowest soliton (quadron) band is gapped with Chern number $C=1$ ($\langle \Delta \hat X\rangle =1$). b) $U=3.0$: lowest two bands merge but (except for finite-size effects) overlap with 
    extended states (unstable quadron) where the transport is
    not quantized. c) $U=4.0$
    lowest two quadron bands merge with vanishing Wilson loop ($\langle \Delta \hat X\rangle =0$). The width of the lines reflects the width of the soliton bands in $K$ space.
   }
    \label{fig:RMM}
\end{figure}

Based on solutions of the mean-field DNLSE it was shown in Ref.~\cite{tuloup2023breakdown} for the example of the Rice-Mele model \cite{rice1982elementary,lohse2016thouless,hayward2018topological} with local attractive interactions that the transition between quantized topological pumping of a soliton to zero pumping upon increasing the interaction strength may go through an intermediate regime of non-quantized transport.
This transient failure of transport quantization, where $\langle \Delta \hat X\rangle$ strongly fluctuates, was attributed in \cite{tuloup2023breakdown}
to a ''self-crossing'' of solutions of the semiclassical DNLSE. 
This phenomenon can be easily understood in the full quantum picture. 
For intermediate interactions, the excited soliton solution is in some parts of the pump cycle degenerate with the continuum of extended states and thus unstable. If this solution merges with the lower soliton band, the time evolution is no longer adiabatic and the transport is not quantized until the interaction is large enough such that also the excited band becomes fully gapped.
In Fig.~\ref{fig:RMM} we show the soliton energies in the Rice-Mele model with different values of the attractive onsite interaction $U$ obtained from exact diagonalization  simulations of a $N=4$ soliton. The width of the lines reflects the width of the soliton bands $E(K,t)$ in $K$ space.
Although only marginally visible due to finite size effects, Fig.~\ref{fig:RMM}b indicates the existence of a parameter range where the lowest soliton band becomes degenerate with an unstable excited solution leading to a fluctuating, non-quantized transport.

\section{Effective Hamiltonian: triplon model}
\label{sec:triplon}

As stated in Sec. \ref{sec:effective_hamiltonian}, we can explicitly construct the effective soliton Hamiltonian  in the strongly interacting limit where the non-interacting part of the Hamiltonian, Eq.~\eqref{eq:aah-hamiltonian}, can be treated as a small perturbation and the solitons become maximally localized (i.e. with a localization length $\xi$ smaller than the lattice spacing). 

In this limit, the binding energy $\frac{U}{2}N(N-1)$ is equal for all soliton solutions and will be disregarded. Transport of the composite object occurs through collective hopping of  particles, arising in $N$th order perturbation theory, which has a very small effective amplitude of $\sim J^N/U^{N-1}$ \cite{Mila_2010}. At the same time, virtual hopping processes of particles from the soliton site to a neighboring site and back give rise to local energy shifts with an amplitude proportional to $\sim J^2/U$.
As shown in detail in the Appendix, this results in the effective energies and hopping amplitudes
\begin{equation}
    \epsilon_{l, \textrm{eff}} = \frac{3}{2} \frac{J_{l-1}^2 + J_l^2}{U},\qquad 
    J_{l,\textrm{eff}} = \frac{3}{2}\frac{J_l^3}{U^2}
    \label{eq:triplon}
\end{equation}
for the triplon model (i.e. $N=3$). (For effective Bloch Hamiltonians of bound doublons and their topological features see also \cite{valiente2008two,PhysRevLett.117.213603,huang2024topological}).
\begin{equation}
    \mathcal{H}_{\textrm{eff}}(t) = - \sum_l \left [ (J_{l, \textrm{eff}}(t)\,  \hat{d}^{\dagger}_{l}\hat{d}_{l+1} + h.c.) + \epsilon_{l, \textrm{eff}}(t) \, \hat{d}^{\dagger}_{l}\hat{d}_{l} \right ].
\end{equation}
%
\begin{figure}[H]
    \centering
    \begin{subfigure}[t]{0.4\textwidth}
        \includegraphics[width=\textwidth]{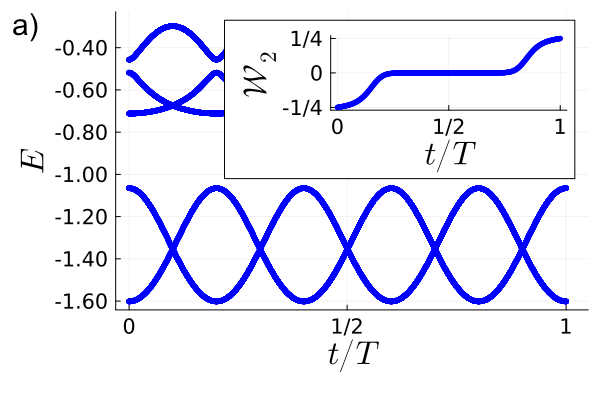}
    \end{subfigure}
    \begin{subfigure}[t]{0.408\textwidth}
        \includegraphics[width=\textwidth]{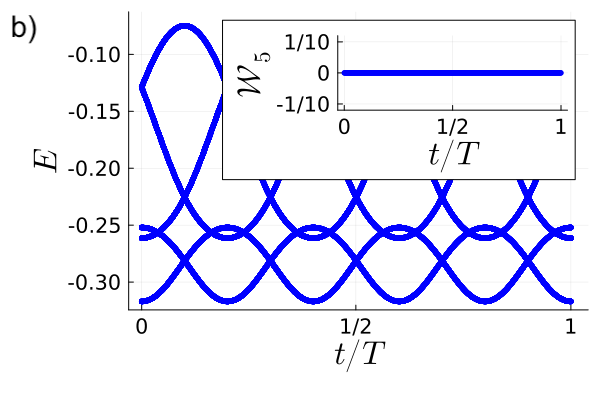}
    \end{subfigure}
    \caption{Instantaneous soliton energies of the effective triplon model, Eq.~\eqref{eq:triplon}, for attractive interaction $U=5$ in (a), and $U=20$ in (b), phase-offset $k=2$ and unit-cell size $p=5$. The insert shows the winding of ${\cal W}=\textrm{Im}\log\det \mathbf{P}(t)$, Eq.~\eqref{eq:wilson-loop}, determining the Wilson loop. 
    }
     \label{fig:triplon}
\end{figure}
%
\noindent

In Fig.~\ref{fig:triplon} we have shown the soliton energies of the effective triplon model for two different interaction strengths for the AAH model along with the integrand of the Wilson-loop in Eq.~\eqref{eq:wilson-loop} ${\cal W}_m=\textrm{Im}\log \det \mathbf{P}(t)$, where $m$ is the number of crossing bands.
One clearly sees that for increased interaction strength the
bands merge. Eventually, all $5$ bands cross and there is no winding of 
${\cal W}_5$, i.e. the 
total Wilson loop vanishes. Thus, despite the fact that the soliton mass is 
still finite, the topological transport vanishes exactly. 

The doublon model ($N=2$) can also be treated analytically and, depending on the lattice model, can  as well show an integer, fractional or trapped case. 
But, since in the doublon case both the effective hopping and the effective potential are  of the same order in interaction strength $U^{-1}$, there is no qualitative change of the bandstructure and therefore no phase-transition.

Furthermore, while the dispersion of wavepackets during a topological pump can be suppressed also for single particles in lattices 
with weakly dispersive (or flat) bands \cite{lee2019}, the interaction-driven topological phase transitions, discussed here, require virtual hopping processes which do not exists in single-particle systems. Thus these transitions are a remarkable and unique feature of composite particles (with more than two particles). 

\begin{figure}[H]
    \centering
    \begin{subfigure}[t]{0.4\textwidth}
        \includegraphics[width=\textwidth]{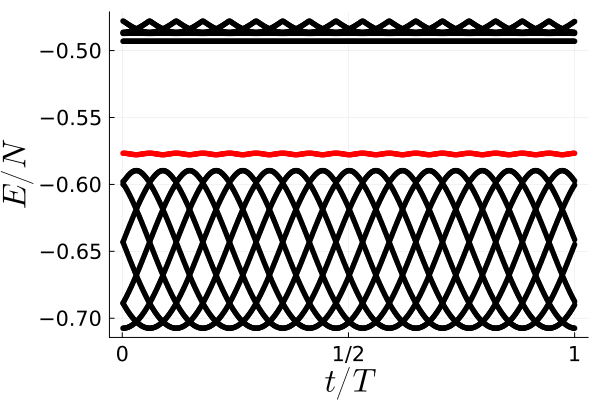}
    \end{subfigure}
    \begin{subfigure}[t]{0.36\textwidth}
        \includegraphics[width=\textwidth]{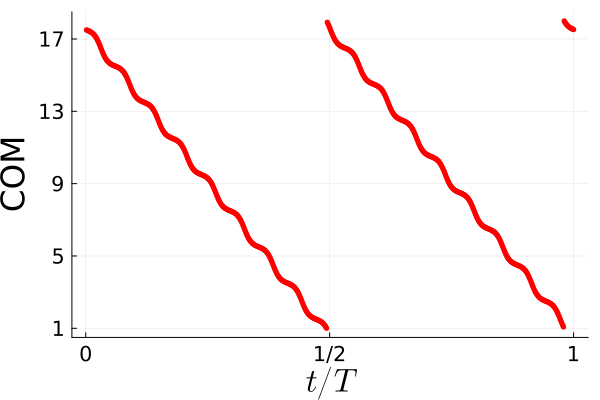}
    \end{subfigure}
    \caption{Instantaneous soliton energies for 3 particles for attractive interaction $U/J=1/3$ with phase-offset $k=8$ and unit-cell size $p=17$ with the corresponding COM-movement in a pump cycle of the red marked energy. The solutions are obtained with the three-site soliton ansatz. 
    }
    \label{fig:excited_solitons}
\end{figure}
\section{Topological transport in higher soliton bands}

So far we mainly discussed the behavior of  ground-state solitons. In single-particle Thouless pumps (like the AAH-model without interaction) quantized topological transport also occurs in higher bands, determined by the corresponding Chern number, where the sum of all Chern-numbers has to vanish. While this is still true in the effective single-particle model, it does not cover the full physics in the composite particle setting. 
Here, excited composite-particle solutions might be unstable (compare Fig. \ref{fig:RMM}), i.e. the soliton  can be energetically degenerate with 
extended, i.e. not self-bound many-particle states.
In such a case the state would not survive in the real time-evolution of the pump because the soliton can evaporate and therefore lose its composite structure.

If, on the other hand, an excited soliton  is stable, there is quantized topological transport determined by the effective Chern number of the excited soliton band, see e.g. Fig.~1a in the extended data of Ref. \cite{jurgensen2021quantized}.
In Fig.~\ref{fig:excited_solitons} we show the energy spectrum from the three-site soliton ansatz of the AAH model for $N=3$ particles with a $p=17$ unit-cell and a $k=8$ phase-offset. The interaction strength is choosen as $U/J=1/3$. As one can see, there are crossing solutions in the low-lying part of the spectrum,  above which another gapped soliton solution exists (marked in red). The corresponding Chern number of this soliton band is $C=2$, which is in agreement with the displayed transport.

\section{summary and outlook}

We developed a full quantum description of topological pumps of self-bound $N$-particle states, i.e. lattice solitons. The quantum description allowed us to identify and to \mbox{explicitly} calculate a topological invariant, i.e. an effective single-particle Chern-number or Wilson-loop explaining the emergence of integer and fractional transport in the full range of interaction strength, where perturbative arguments \cite{jurgensen2022chern,mostaan2022quantized} fail. Transitions between phases of differently quantized topological transport observed in experiments \cite{jurgensen2021quantized,jurgensen2023quantized} 
as well as the possibility of parameter regimes without quantized transport were explained by coalescence of soliton bands and possible degeneracies with extended states. 
Quantized topological transport can also be observed in excited soliton bands, provided these bands are gapped and stable, i.e. there are no degeneracies with extended many-particle states.
The concept can easily be extended to multi-component solitons \cite{PhysRevResearch.6.L042010,tao2024nonlinearity,tao2025nonlinearity}, were, among other things, fractional transport was predicted despite the fact that all single-particle bands are topologically trivial. In the latter case interactions not only lead to the modification of topological properties, such as the transition from integer to fractional phases, but to the \textit{emergence} of topology. 
Finally, we note that the concept can also be applied to bound many-particle states in cold gas experiments and topological pumps \cite{valiente2008two,winkler2006repulsively,bertok2022splitting,walter2023quantization,viebahn2024interactions}.

\paragraph*{Acknowledgements}

The authors thank Alaa Bayazeed for fruitful discussions. Financial support from the DFG through SFB TR 185, Project No. 277625399, is gratefully acknowledged. The authors also thank the Allianz für Hochleistungsrechnen (AHRP) for giving us access to the “Elwetritsch” HPC Cluster. Some of the data in this script has been obtained using the \textit{QuantumOptics.jl} framework \cite{kramer2018quantumoptics}.

\section*{Appendix}

\subsection*{Effective soliton Hamiltonians in the limit of strongly localized solitons}
\label{sec:effective_methods}

To derive an effective model for strongly localized composite particles, let us first look at the binding energy of such a maximally localized soliton in the AAH model \eqref{eq:aah-hamiltonian} (i.e. a localization length $\xi$ smaller than the lattice spacing) which has a binding energy given by 
 \begin{align}
     E_{\textrm{int}, N} = \frac{U}{2} N (N-1)
 \end{align}
 whereas states with $N-1$ localized particles have an energy
 \begin{align}
     E_{\textrm{int}, N-1} = \frac{U}{2} (N-1) (N-2).
 \end{align}
The energy difference $\vert E_{\textrm{int}, N} - E_{\textrm{int}, N-1}\vert$ is $U(N-1)$. So if $U$ is sufficiently larger than the bose-enhanced hopping $\sqrt{N} J$, the energetically lowest states are localized solitons.

To derive an effective description of the maximally localized solitons, as proposed in Eq. \eqref{eq:effective_hamiltonian}, we have to identify the relevant processes, treating the particle hopping as perturbation. These processes have to be resonant between the degenerate groundstates of the interaction Hamiltonian $\mathcal{H}_{\textrm{int}}$ \cite{Mila_2010}:
For the effective potential, it is the virtual hopping of a single-particle occuring in second order perturbation theory
\begin{align*}
  \sim  \hat{a}^{\dagger}_{l}\hat{a}_{l+1} \hat{a}^{\dagger}_{l+1}\hat{a}_{l} \vert \Psi \rangle.
\end{align*}
On the other hand, the effective hopping of the complete composite particle emerges only in $N$th order perturbation theory 
\begin{align*}
    \sim (\hat{a}^{\dagger}_{l+1} \hat{a}_{l})^{N} \vert \Psi \rangle.
\end{align*}
The amplitude for the virtual hopping scales as $ \sim 1/U$, while the effective hopping goes with $ \sim 1/U^{N-1}$.

\paragraph*{The minimal particle number: Triplons --} In the following we will explicitly calculate an effective Hamiltonian for the minimal possible particle number: the triplon. 
For a composite object consisting of only two particles,  both processes (virtual and effective hopping) would have the same scaling $1/U$ and therefore the interaction has no qualitative influence on the system properties. 

The effective potential is calculated directly within \mbox{second} order perturbation theory of the Hamiltonian and is given as
\begin{equation}
    \epsilon_{l, \textrm{eff}} = \frac{3}{2} \frac{J_{l-1}^2 + J_l^2}{U}.
    \label{eq:effective_potential_triplon}
\end{equation}
Calculating the effective triplon hopping in perturbation theory is possible but already requires good bookkeeping since it is a third order process. 

Therefore, we will take a look at the local basis of two sites  for three particles:
\begin{equation*}
    \vert 30 \rangle, \vert 21 \rangle, \vert 12 \rangle, \vert 03 \rangle.
\end{equation*}
The local Hamiltonian for these states 
can be written in matrix form:
\begin{equation*}
    \left (
    \begin{matrix}
        -3U & -\sqrt{3}J_l & 0 & 0 \\
        -\sqrt{3}J_l & -U & -2J_l & 0 \\
        0 & -2J_l & -U & -\sqrt{3}J_l \\
        0 & 0 & -\sqrt{3}J_l & -3U 
    \end{matrix}
    \right ) .
\end{equation*}
Here we assume - without loss of generality - the left site is located at position $l$ in our system. Diagonalizing this 4x4-matrix yields four eigenstates. The two low-energy eigenstates are (for sufficient large values of the interaction strength $U$):
\begin{equation*}
    \vert\psi_\pm\rangle \propto \vert 30 \rangle \pm \vert 03 \rangle.
\end{equation*}
The corresponding eigenenergies are:
\begin{equation*}
    E_\pm \propto \mp t - 2U - \sqrt{4t^2 \mp 2tU + U^2}.
\end{equation*}
In the effective Hamiltonian (compare Eq.~\eqref{eq:effective_hamiltonian}) the eigenenergies of $\vert\psi_\pm\rangle$ can be shown to be
\begin{eqnarray*}
    E_{\textrm{eff}, \pm} &&= \mp J_{l,\textrm{eff}} + \textrm{const} \\
    E_{\textrm{eff}, +} - E_{\textrm{eff}, -}  && = -2 J_{l,\textrm{eff}}
\end{eqnarray*}
Given these two relations we can Taylor expand the energy difference $E_+ - E_-$ for small values of the hopping amplitude and extract the effective hopping:
\begin{equation}
    J_{l,\textrm{eff}} = -\frac{E_+ - E_-}{2} = \frac{3}{2}\frac{J_l^3}{U^2} + \mathcal{O}\left(\frac{J_l^5}{U^4}\right).
    \label{eq:effective_hopping_triplon}
\end{equation}
With both the virtual and the effective hopping Eqs.~\eqref{eq:effective_potential_triplon} and \eqref{eq:effective_hopping_triplon} we can calculate an effective single-particle Hamiltonian Eq.~\eqref{eq:effective_hamiltonian} reflecting the same physics as the full model in the strong interacting limit.


\bibliography{soliton-topology.bib}

@article{aidelsburger_review_thouless,
    author = {Citro, Roberta and Aidelsburger, Monika},
    title = {Thouless pumping and topology},
    journal = {Nat Rev Phys},
    volume = {5},
    pages = {87-101},
    year = {2023},
    url = {https://doi.org/10.1038/s42254-022-00545-0}
}

@article{thouless_pump,
  title = {Quantization of particle transport},
  author = {Thouless, D. J.},
  journal = {Phys. Rev. B},
  volume = {27},
  issue = {10},
  pages = {6083--6087},
  numpages = {0},
  year = {1983},
  month = {May},
  publisher = {American Physical Society},
  doi = {10.1103/PhysRevB.27.6083},
  url = {https://link.aps.org/doi/10.1103/PhysRevB.27.6083}
}

@inbook{Mila_2010,
   title={Strong-Coupling Expansion and Effective Hamiltonians},
   ISBN={9783642105890},
   ISSN={0171-1873},
   url={http://dx.doi.org/10.1007/978-3-642-10589-0_20},
   DOI={10.1007/978-3-642-10589-0_20},
   booktitle={Introduction to Frustrated Magnetism},
   publisher={Springer Berlin Heidelberg},
   author={Mila, Frédéric and Schmidt, Kai Phillip},
   year={2010},
   month=sep, pages={537–559} 
}

@article{niu_chern_number,
  title = {Quantized Hall conductance as a topological invariant},
  author = {Niu, Qian and Thouless, D. J. and Wu, Yong-Shi},
  journal = {Phys. Rev. B},
  volume = {31},
  issue = {6},
  pages = {3372--3377},
  numpages = {0},
  year = {1985},
  month = {Mar},
  publisher = {American Physical Society},
  doi = {10.1103/PhysRevB.31.3372},
  url = {https://link.aps.org/doi/10.1103/PhysRevB.31.3372}
}

@article{wilczek-zee,
  title = {Appearance of Gauge Structure in Simple Dynamical Systems},
  author = {Wilczek, Frank and Zee, A.},
  journal = {Phys. Rev. Lett.},
  volume = {52},
  issue = {24},
  pages = {2111--2114},
  numpages = {0},
  year = {1984},
  month = {Jun},
  publisher = {American Physical Society},
  doi = {10.1103/PhysRevLett.52.2111},
  url = {https://link.aps.org/doi/10.1103/PhysRevLett.52.2111}
}

@article{klitzing1980new,
  title = {New Method for High-Accuracy Determination of the Fine-Structure Constant Based on Quantized Hall Resistance},
  author = {Klitzing, K. v. and Dorda, G. and Pepper, M.},
  journal = {Phys. Rev. Lett.},
  volume = {45},
  issue = {6},
  pages = {494--497},
  numpages = {0},
  year = {1980},
  month = {Aug},
  publisher = {American Physical Society},
  doi = {10.1103/PhysRevLett.45.494},
  url = {https://link.aps.org/doi/10.1103/PhysRevLett.45.494}
}

@article{thouless1982quantized,
  title = {Quantized Hall Conductance in a Two-Dimensional Periodic Potential},
  author = {Thouless, D. J. and Kohmoto, M. and Nightingale, M. P. and den Nijs, M.},
  journal = {Phys. Rev. Lett.},
  volume = {49},
  issue = {6},
  pages = {405--408},
  numpages = {0},
  year = {1982},
  month = {Aug},
  publisher = {American Physical Society},
  doi = {10.1103/PhysRevLett.49.405},
  url = {https://link.aps.org/doi/10.1103/PhysRevLett.49.405}
}

@article{altland1997nonstandard,
  title = {Nonstandard symmetry classes in mesoscopic normal-superconducting hybrid structures},
  author = {Altland, Alexander and Zirnbauer, Martin R.},
  journal = {Phys. Rev. B},
  volume = {55},
  issue = {2},
  pages = {1142--1161},
  numpages = {0},
  year = {1997},
  month = {Jan},
  publisher = {American Physical Society},
  doi = {10.1103/PhysRevB.55.1142},
  url = {https://link.aps.org/doi/10.1103/PhysRevB.55.1142}
}

@article{lederer2008discrete,
title = {Discrete solitons in optics},
journal = {Physics Reports},
volume = {463},
number = {1},
pages = {1-126},
year = {2008},
issn = {0370-1573},
doi = {https://doi.org/10.1016/j.physrep.2008.04.004},
url = {https://www.sciencedirect.com/science/article/pii/S0370157308001257},
author = {Falk Lederer and George I. Stegeman and Demetri N. Christodoulides and Gaetano Assanto and Moti Segev and Yaron Silberberg}
}

@article{jurgensen2021quantized,
  title={Quantized nonlinear Thouless pumping},
  author={J{\"u}rgensen, Marius and Mukherjee, Sebabrata and Rechtsman, Mikael C},
  journal={Nature},
  volume={596},
  number={7870},
  pages={63--67},
  year={2021},
  publisher={Nature Publishing Group UK London},
    url = {https://doi.org/10.1038/s41586-021-03688-9}
}

@article{jurgensen2023quantized,
  title={Quantized fractional Thouless pumping of solitons},
  author={J{\"u}rgensen, Marius and Mukherjee, Sebabrata and J{\"o}rg, Christina and Rechtsman, Mikael C},
  journal={Nature Physics},
  volume={19},
  number={3},
  pages={420--426},
  year={2023},
  publisher={Nature Publishing Group UK London},
    url = {https://www.nature.com/articles/s41567-022-01871-x}
}

@article{jurgensen2022chern,
  title = {Chern Number Governs Soliton Motion in Nonlinear Thouless Pumps},
  author = {J\"urgensen, Marius and Rechtsman, Mikael C.},
  journal = {Phys. Rev. Lett.},
  volume = {128},
  issue = {11},
  pages = {113901},
  numpages = {6},
  year = {2022},
  month = {Mar},
  publisher = {American Physical Society},
  doi = {10.1103/PhysRevLett.128.113901},
  url = {https://link.aps.org/doi/10.1103/PhysRevLett.128.113901}
}

@article{mostaan2022quantized,
  title={Quantized topological pumping of solitons in nonlinear photonics and ultracold atomic mixtures},
  author={Mostaan, Nader and Grusdt, Fabian and Goldman, Nathan},
  journal={Nature Communications},
  volume={13},
  number={1},
  pages={5997},
  year={2022},
  publisher={Nature Publishing Group UK London},
    url = {https://www.nature.com/articles/s41467-022-33478-4}
}

@article{scott1994quantum,
title = {Quantum lattice solitons},
journal = {Physica D: Nonlinear Phenomena},
volume = {78},
number = {3},
pages = {194-213},
year = {1994},
issn = {0167-2789},
doi = {https://doi.org/10.1016/0167-2789(94)90115-5},
url = {https://www.sciencedirect.com/science/article/pii/0167278994901155},
author = {A.C. Scott and J.C. Eilbeck and H. Gilhøj}
}

@article{tuloup2023breakdown,
  title={Breakdown of quantization in nonlinear Thouless pumping},
  author={Tuloup, Thomas and Bomantara, Raditya Weda and Gong, Jiangbin},
  journal={New J. Phys.},
  volume={25},
  number={8},
  pages={083048},
  year={2023},
  publisher={IOP Publishing},
    url = {https://iopscience.iop.org/article/10.1088/1367-2630/acef4d}
}

@article{hu2024pumping,
  title={Pumping of matter wave solitons in one-dimensional optical superlattices},
  author={Hu, Xiaoxiao and Li, Zhiqiang and Chen, Ai-Xi and Luo, Xiaobing},
  journal={New J. Phys.},
  volume={26},
  number={12},
  pages={123006},
  year={2024},
  publisher={IOP Publishing},
    url = {https://iopscience.iop.org/article/10.1088/1367-2630/ad9770}
}

@article{mattis1986few,
  title = {The few-body problem on a lattice},
  author = {Mattis, Daniel C.},
  journal = {Rev. Mod. Phys.},
  volume = {58},
  issue = {2},
  pages = {361--379},
  numpages = {0},
  year = {1986},
  month = {Apr},
  publisher = {American Physical Society},
  doi = {10.1103/RevModPhys.58.361},
  url = {https://link.aps.org/doi/10.1103/RevModPhys.58.361}
}

@article{valiente2008two,
  title={Two-particle states in the Hubbard model},
  author={Valiente, Manuel and Petrosyan, David},
  journal={ J. Phys. B: At. Mol. Opt. Phys.},
  volume={41},
  number={16},
  pages={161002},
  year={2008},
  publisher={IOP Publishing},
    url = {https://iopscience.iop.org/article/10.1088/0953-4075/41/16/161002}
}

@article{naldesi2019rise,
  title = {Rise and Fall of a Bright Soliton in an Optical Lattice},
  author = {Naldesi, Piero and Gomez, Juan Polo and Malomed, Boris and Olshanii, Maxim and Minguzzi, Anna and Amico, Luigi},
  journal = {Phys. Rev. Lett.},
  volume = {122},
  issue = {5},
  pages = {053001},
  numpages = {6},
  year = {2019},
  month = {Feb},
  publisher = {American Physical Society},
  doi = {10.1103/PhysRevLett.122.053001},
  url = {https://link.aps.org/doi/10.1103/PhysRevLett.122.053001}
}

@article{barbiero2014quantum,
  title = {Quantum bright solitons in a quasi-one-dimensional optical lattice},
  author = {Barbiero, Luca and Salasnich, Luca},
  journal = {Phys. Rev. A},
  volume = {89},
  issue = {6},
  pages = {063605},
  numpages = {5},
  year = {2014},
  month = {Jun},
  publisher = {American Physical Society},
  doi = {10.1103/PhysRevA.89.063605},
  url = {https://link.aps.org/doi/10.1103/PhysRevA.89.063605}
}

@book{korepin1997quantum,
  title={Quantum inverse scattering method and correlation functions},
  author={Korepin, Vladimir E and Korepin, Vladimir E and Bogoliubov, NM and Izergin, AG},
  volume={3},
  year={1997},
  publisher={Cambridge university press},
    url = {https://www.cambridge.org/core/books/quantum-inverse-scattering-method-and-correlation-functions/CF36D7B224AC61B8D67678D14E92C64F}
}

@article{aubry1980analyticity,
  title={Analyticity breaking and Anderson localization in incommensurate lattices},
  author={Aubry, Serge and Andr{\'e}, Gilles},
  journal={Ann. Israel Phys. Soc},
  volume={3},
  number={133},
  pages={18},
  year={1980},
    url = {https://chaos.if.uj.edu.pl/~delande/Lectures/files/An.Is.Phys.Soc.pdf}
}

@article{harper1955single,
  title={Single band motion of conduction electrons in a uniform magnetic field},
  author={Harper, Philip George},
  journal={Proc. Phys. Soc. A},
  volume={68},
  number={10},
  pages={874},
  year={1955},
  publisher={IOP Publishing},
    url = {https://iopscience.iop.org/article/10.1088/0370-1298/68/10/304}
}

@misc{tao2025nonlinearity,
      title={Nonlinearity-induced Fractional Thouless Pumping of Solitons}, 
      author={Yu-Liang Tao and Yongping Zhang and Yong Xu},
      year={2025},
      eprint={2502.06131},
      archivePrefix={arXiv},
      url={https://arxiv.org/abs/2502.06131}, 
}

@article{rice1982elementary,
  title = {Elementary Excitations of a Linearly Conjugated Diatomic Polymer},
  author = {Rice, M. J. and Mele, E. J.},
  journal = {Phys. Rev. Lett.},
  volume = {49},
  issue = {19},
  pages = {1455--1459},
  numpages = {0},
  year = {1982},
  month = {Nov},
  publisher = {American Physical Society},
  doi = {10.1103/PhysRevLett.49.1455},
  url = {https://link.aps.org/doi/10.1103/PhysRevLett.49.1455}
}

@article{hayward2018topological,
  title = {Topological charge pumping in the interacting bosonic Rice-Mele model},
  author = {Hayward, A. and Schweizer, C. and Lohse, M. and Aidelsburger, M. and Heidrich-Meisner, F.},
  journal = {Phys. Rev. B},
  volume = {98},
  issue = {24},
  pages = {245148},
  numpages = {11},
  year = {2018},
  month = {Dec},
  publisher = {American Physical Society},
  doi = {10.1103/PhysRevB.98.245148},
  url = {https://link.aps.org/doi/10.1103/PhysRevB.98.245148}
}

@article{lohse2016thouless,
  title={A Thouless quantum pump with ultracold bosonic atoms in an optical superlattice},
  author={Lohse, Michael and Schweizer, Christian and Zilberberg, Oded and Aidelsburger, Monika and Bloch, Immanuel},
  journal={Nature Physics},
  volume={12},
  number={4},
  pages={350--354},
  year={2016},
  publisher={Nature Publishing Group UK London},
    url = {https://www.nature.com/articles/nphys3584}
}

@article{fu2022nonlinear,
  title = {Nonlinear Thouless Pumping: Solitons and Transport Breakdown},
  author = {Fu, Qidong and Wang, Peng and Kartashov, Yaroslav V. and Konotop, Vladimir V. and Ye, Fangwei},
  journal = {Phys. Rev. Lett.},
  volume = {128},
  issue = {15},
  pages = {154101},
  numpages = {6},
  year = {2022},
  month = {Apr},
  publisher = {American Physical Society},
  doi = {10.1103/PhysRevLett.128.154101},
  url = {https://link.aps.org/doi/10.1103/PhysRevLett.128.154101}
}

@article{kraus2012topological,
  title = {Topological States and Adiabatic Pumping in Quasicrystals},
  author = {Kraus, Yaacov E. and Lahini, Yoav and Ringel, Zohar and Verbin, Mor and Zilberberg, Oded},
  journal = {Phys. Rev. Lett.},
  volume = {109},
  issue = {10},
  pages = {106402},
  numpages = {5},
  year = {2012},
  month = {Sep},
  publisher = {American Physical Society},
  doi = {10.1103/PhysRevLett.109.106402},
  url = {https://link.aps.org/doi/10.1103/PhysRevLett.109.106402}
}

@misc{tao2024nonlinearity,
      title={Nonlinearity-induced Thouless pumping of solitons}, 
      author={Yu-Liang Tao and Jiong-Hao Wang and Yong Xu},
      year={2024},
      eprint={2409.19515},
      archivePrefix={arXiv},
      url={https://arxiv.org/abs/2409.19515}, 
}

@article{lumer2013self,
  title = {Self-Localized States in Photonic Topological Insulators},
  author = {Lumer, Yaakov and Plotnik, Yonatan and Rechtsman, Mikael C. and Segev, Mordechai},
  journal = {Phys. Rev. Lett.},
  volume = {111},
  issue = {24},
  pages = {243905},
  numpages = {5},
  year = {2013},
  month = {Dec},
  publisher = {American Physical Society},
  doi = {10.1103/PhysRevLett.111.243905},
  url = {https://link.aps.org/doi/10.1103/PhysRevLett.111.243905}
}

@article{mukherjee2020observation,
  title={Observation of Floquet solitons in a topological bandgap},
  author={Mukherjee, Sebabrata and Rechtsman, Mikael C},
  journal={Science},
  volume={368},
  number={6493},
  pages={856--859},
  year={2020},
  publisher={American Association for the Advancement of Science},
    url = {https://www.science.org/doi/10.1126/science.aba8725}
}

@article{PhysRevLett.117.213603,
  title = {Topological Pumping of Photons in Nonlinear Resonator Arrays},
  author = {Tangpanitanon, Jirawat and Bastidas, Victor M. and Al-Assam, Sarah and Roushan, Pedram and Jaksch, Dieter and Angelakis, Dimitris G.},
  journal = {Phys. Rev. Lett.},
  volume = {117},
  issue = {21},
  pages = {213603},
  numpages = {5},
  year = {2016},
  month = {Nov},
  publisher = {American Physical Society},
  doi = {10.1103/PhysRevLett.117.213603},
  url = {https://link.aps.org/doi/10.1103/PhysRevLett.117.213603}
}

@article{ke2017multiparticle,
  title = {Multiparticle Wannier states and Thouless pumping of interacting bosons},
  author = {Ke, Yongguan and Qin, Xizhou and Kivshar, Yuri S. and Lee, Chaohong},
  journal = {Phys. Rev. A},
  volume = {95},
  issue = {6},
  pages = {063630},
  numpages = {11},
  year = {2017},
  month = {Jun},
  publisher = {American Physical Society},
  doi = {10.1103/PhysRevA.95.063630},
  url = {https://link.aps.org/doi/10.1103/PhysRevA.95.063630}
}

@article{gutzwiller,
  title = {Effect of Correlation on the Ferromagnetism of Transition Metals},
  author = {Gutzwiller, Martin C.},
  journal = {Phys. Rev. Lett.},
  volume = {10},
  issue = {5},
  pages = {159--162},
  numpages = {0},
  year = {1963},
  month = {Mar},
  publisher = {American Physical Society},
  doi = {10.1103/PhysRevLett.10.159},
  url = {https://link.aps.org/doi/10.1103/PhysRevLett.10.159}
}

@article{huang2024topological,
  title={Topological pumping induced by spatiotemporal modulation of interaction},
  author={Huang, Boning and Ke, Yongguan and Liu, Wenjie and Lee, Chaohong},
  journal={Phys. Scr.},
  volume={99},
  number={6},
  pages={065997},
  year={2024},
  publisher={IOP Publishing},
url={https://iopscience.iop.org/article/10.1088/1402-4896/ad491e}
}

@article{PhysRevResearch.6.L042010,
  title = {Thouless pumping and trapping of two-component gap solitons},
  author = {Lyu, Hao and Zhang, Yongping and Busch, Thomas},
  journal = {Phys. Rev. Res.},
  volume = {6},
  issue = {4},
  pages = {L042010},
  numpages = {7},
  year = {2024},
  month = {Oct},
  publisher = {American Physical Society},
  doi = {10.1103/PhysRevResearch.6.L042010},
  url = {https://link.aps.org/doi/10.1103/PhysRevResearch.6.L042010}
}

@article{ryu2010topological,
  title={Topological insulators and superconductors: tenfold way and dimensional hierarchy},
  author={Ryu, Shinsei and Schnyder, Andreas P and Furusaki, Akira and Ludwig, Andreas WW},
  journal={New Journal of Physics},
  volume={12},
  number={6},
  pages={065010},
  year={2010},
  publisher={IOP Publishing},
  url={https://doi.org/10.1088/1367-2630/12/6/065010}
}

@article{sone2024nonlinearity,
  title={Nonlinearity-induced topological phase transition characterized by the nonlinear Chern number},
  author={Sone, Kazuki and Ezawa, Motohiko and Ashida, Yuto and Yoshioka, Nobuyuki and Sagawa, Takahiro},
  journal={Nature Physics},
  volume={20},
  number={7},
  pages={1164--1170},
  year={2024},
  publisher={Nature Publishing Group UK London},
  url={https://doi.org/10.1038/s41567-024-02451-x}
}

@article{kramer2018quantumoptics,
  title={QuantumOptics. jl: A Julia framework for simulating open quantum systems},
  author={Kr{\"a}mer, Sebastian and Plankensteiner, David and Ostermann, Laurin and Ritsch, Helmut},
  journal={Computer Physics Communications},
  volume={227},
  pages={109--116},
  year={2018},
  publisher={Elsevier},
  url={https://www.sciencedirect.com/science/article/pii/S0010465518300328}
}

@misc{jurgensen2025fermions,
      title={Multi-band fractional Thouless pumps}, 
      author={Marius Jürgensen and Jacob Steiner and Gil Refael and Mikael C. Rechtsman},
      year={2025},
      eprint={2504.09338},
      archivePrefix={arXiv},
      url={https://arxiv.org/abs/2504.09338}
}

@article{sone2025transition,
  title={Transition from the topological to the chaotic in the nonlinear Su--Schrieffer--Heeger model},
  author={Sone, Kazuki and Ezawa, Motohiko and Gong, Zongping and Sawada, Taro and Yoshioka, Nobuyuki and Sagawa, Takahiro},
  journal={Nature Communications},
  volume={16},
  number={1},
  pages={422},
  year={2025},
  publisher={Nature Publishing Group UK London},
  url={https://doi.org/10.1038/s41467-024-55237-3}
}

@article{zhou2022topological,
  title={Topological invariant and anomalous edge modes of strongly nonlinear systems},
  author={Zhou, Di and Rocklin, D Zeb and Leamy, Michael and Yao, Yugui},
  journal={Nature Communications},
  volume={13},
  number={1},
  pages={3379},
  year={2022},
  publisher={Nature Publishing Group UK London},
  url={https://doi.org/10.1038/s41467-022-31084-y}
}

@article{lee2019,
  title = {Dispersion-suppressed topological Thouless pumping},
  author = {Hu, Shi and Ke, Yongguan and Deng, Yuangang and Lee, Chaohong},
  journal = {Phys. Rev. B},
  volume = {100},
  issue = {6},
  pages = {064302},
  numpages = {9},
  year = {2019},
  month = {Aug},
  publisher = {American Physical Society},
  doi = {10.1103/PhysRevB.100.064302},
  url = {https://link.aps.org/doi/10.1103/PhysRevB.100.064302}
}

@article{winkler2006repulsively,
  title={Repulsively bound atom pairs in an optical lattice},
  author={Winkler, K and Thalhammer, G and Lang, F and Grimm, R and Hecker Denschlag, J and Daley, AJ and Kantian, A and B{\"u}chler, HP and Zoller, P},
  journal={Nature},
  volume={441},
  number={7095},
  pages={853--856},
  year={2006},
  publisher={Nature Publishing Group UK London},
  url={https://doi.org/10.1038/nature04918}
}

@article{walter2023quantization,
  title={Quantization and its breakdown in a Hubbard--Thouless pump},
  author={Walter, Anne-Sophie and Zhu, Zijie and G{\"a}chter, Marius and Minguzzi, Joaqu{\'\i}n and Roschinski, Stephan and Sandholzer, Kilian and Viebahn, Konrad and Esslinger, Tilman},
  journal={Nature Physics},
  volume={19},
  number={10},
  pages={1471--1475},
  year={2023},
  publisher={Nature Publishing Group UK London},
  url={https://doi.org/10.1038/s41567-023-02145-w}
}

@article{viebahn2024interactions,
  title={Interactions enable Thouless pumping in a nonsliding lattice},
  author={Viebahn, Konrad and Walter, Anne-Sophie and Bertok, Eric and Zhu, Zijie and G{\"a}chter, Marius and Aligia, Armando A and Heidrich-Meisner, Fabian and Esslinger, Tilman},
  journal={Physical Review X},
  volume={14},
  number={2},
  pages={021049},
  year={2024},
  publisher={APS},
  url={https://doi.org/10.1103/PhysRevX.14.021049}
}

@article{bertok2022splitting,
  title={Splitting of topological charge pumping in an interacting two-component fermionic Rice-Mele Hubbard model},
  author={Bertok, Eric and Heidrich-Meisner, Fabian and Aligia, AA},
  journal={Physical Review B},
  volume={106},
  number={4},
  pages={045141},
  year={2022},
  publisher={APS},
  url={https://doi.org/10.1103/PhysRevB.106.045141}
}

\end{document}